# Rejoinder: Bayesian Checking of the Second Levels of Hierarchical Models

**M. J. Bayarri and M. E. Castellanos**

We would like to thank the discussants for the valuable insights and for commenting on important aspects of model checking that we did not touch in our paper. Our goal was modest (but crucial): to select an appropriate distribution with which to judge the compatibility of the data with a hypothesized (hierarchical) model, when the test statistic is not ancillary and an improper prior is used for the hyperparameters. Since it is important to emphasize that this is by no means the only aspect of model checking, the discussants' complementary contributions and comments are all most welcome. The specific technical contributions of Evans and Johnson are also appreciated, since their developments in this area were not mentioned in our review.

Several discussants have highlighted the importance of graphical displays in model checking. We will not comment on this because we entirely agree. We similarly agree with most of the discussants' other comments, although in this rejoinder we mainly concentrate on disagreements. Our comments are organized around the main topics that arise in the discussions. We keep the same notation and terminology used in the paper (although it does conflict with the notation used by some of the discussants).

*M. J. Bayarri is Professor, Department of Statistics and Operation Research, University of Valencia, Burjassot, (Valencia), 46100 Spain e-mail: susie.bayarri@uv.es. M. E. Castellanos is Associate Professor, Department of Statistics and Operation Research, Rey Juan Carlos University, Móstoles, (Madrid), 28933 Spain e-mail: maria.castellanos@urjc.es.*

## ROLE OF PRIOR PREDICTIVE DISTRIBUTIONS WHEN MODEL UNCERTAINTY IS PRESENT

Bayesian analyses, when model uncertainty is present (model choice, model averaging), are based on the prior predictive distributions for the different models under consideration. Model checking is a quick-and-dirty shortcut to bypass model choice, and "pure" Bayesian reasoning indicates that all relevant information lies in the (prior) predictive distribution $m(\mathbf{x})$ for the entertained model.

As Evans points out, objective Bayes methodology should be guided by proper Bayes methodology, so objective Bayes model checking should also be based on the prior predictive distribution. The difficulty, however, is that only some aspects of this distribution can be utilized when the prior distribution is improper. Bayarri and Berger (1997, 1999, 2000) argue that the relevant aspect to consider for model checking is a conditional (prior) predictive distribution $m(\mathbf{x} \mid u)$, where $U = U(\mathbf{X})$ is an appropriate conditioning statistic such that the posterior $\pi(\boldsymbol{\theta} \mid u)$ is proper. Model checks (measures of surprise) computed with this distribution (such as $p$-values or relative surprise) are called *conditional predictive* measures.

If we use a statistic $T$ to measure departure and use $U$ for conditioning, the relevant distribution for model checking is then $m(t \mid u)$. Evans' prescription can be put in this framework with $T$ ancillary and $U$ sufficient (caution: Evans' notation switches the roles of $T$ and $U$). Larsen and Lu's (from now on L&L) prescription for checking group $i$ is also of this form with $T = T(\mathbf{X}_i)$ and $U = \mathbf{X}_{(-i)}$. The complete theory of Johnson (not sketched in his discussion) relies on the whole prior predictive. Hence, all these methods produce legitimate Bayesian measures of surprise. The posterior predictive distribution cannot be expressed in this way (it would produce a trivial, degenerate distribution).

Bayarri and Berger (1997, 1999) explore several choices of $U$ and recommend use of the *conditional*

MLE of $\boldsymbol{\theta}$, that is, the MLE computed in the conditional distribution $f(\mathbf{x} \mid t, \boldsymbol{\theta})$. The resulting measures of surprise (or model checks) were shown to basically coincide with the partial posterior measures; indeed, the conditional predictive distribution for that choice of $U$ and the partial posterior predictive distribution are asymptotically equivalent (Robins (1999); Robins, van der Vaart and Ventura (2000)).

We have concentrated on partial posterior measures because they are basically indistinguishable from the conditional predictive ones and they are easier to compute, but their Bayesian justification comes from the conditional predictive reasoning. We should perhaps have reiterated this in the paper.

## CHOICE OF $T$ AND/OR $D$

We are not addressing optimal choice of $T$ in this paper: we focus on the choice of the relevant distribution to locate $T$. $T$ is often chosen casually based on intuitive grounds and we wanted a method that would work with *any* choice of the departure statistic $T$ (although, of course, adequate choice of $T$ is always important to increase power). However, several discussants have focused their discussion on specific choices, so we comment on those.

A preliminary issue is consideration of *discrepancy measures*, that is, functions of the data and the parameters $D = D(\mathbf{x}, \boldsymbol{\theta})$, as well as statistics $T = T(\mathbf{x})$ for model checking. Gelman and L&L favor their routine use, also with informal, intuitively sound choices. Johnson's proposal, although derived from a different philosophy, could also be considered under this umbrella. Johnson's interesting method applies to invariant situations in which the distribution of an optimally chosen $D$, namely a pivotal quantity, is precisely known. Johnson's elegant theorem shows how to obtain simulations from the pivotal quantities for the true (unknown) parameter values, so that their adequacy with the known distribution can be assessed. The main difficulty is that these simulations are highly correlated and proper assessments require prior predictive techniques (and hence informative priors). In some situations, the provided bounds for the $p$-values of the suggested test statistic might suffice, so these techniques are definitely worth considering. Note, however, that without an informative prior, interpretation of graphical displays, or other uses of these correlated simulations, is an issue.

Although our methodology could be applied to such functions [it would probably suffice to consider the joint conditional distribution $p(\mathbf{x}, \boldsymbol{\theta} \mid u)$], we have not thought about it enough to venture an opinion. Use of $D$'s seems intuitive; however, when used in conjunction with posterior predictive distributions, they suffer from the same type of conservativeness as statistics do (Robins (1999); Robins, van der Vaart and Ventura (2000)). Since the problems are the same whether or not $T$ is chosen to also include parameters, we cast the rest of the rejoinder in terms of traditional statistics $T$. (Note that, if $T$ is ancillary or $D$ pivotal, the issues about how to integrate out the parameters disappear.)

Evans chooses not to integrate out the unknown $\boldsymbol{\theta}$ but rather to eliminate it in traditional frequentist ways, by either conditioning on a sufficient statistic (i.e., $U$ above is sufficient) or by using an ancillary test statistic $T$. His argument is, however, also well within Bayesian thinking, providing a beautiful factorization of the joint (prior) distribution of $\mathbf{x}$ and $\boldsymbol{\theta}$ in which the role of the different factors can be very nicely interpreted. Although these specific choices of $T$ and $U$ are needed for the clean factorization, we show that other choices of $T$ and/or $U$ are also possible (maybe desirable) for model checking, and might be simpler to implement. This applies specially to problems in which the required statistics do not exist, are difficult to identify, or when sampling from the resulting distribution is particularly challenging.

Johnson wonders about choices of $T$ sufficient (or nearly so) and/or $T$ ancillary. $T$ should *not* be sufficient; a sufficient $T$ is virtually useless for model checking (this is in agreement with Evans' remarks). An extensive discussion of this issue, with examples, can be found in Bayarri and Berger (1997), Bayarri and Berger (2000) and rejoinder. An ancillary $T$ simply reproduces frequentist testing with *similar* $p$-values (terminology from Bayarri and Berger (1999), 2000); the Bayesian machinery for integrating out unknown quantities is simply not needed and, in this case, prior, posterior, conditional and partial posterior predictive distributions are all identical to the specified marginal distribution for $T$, $f(t)$. When $T$ is nearly ancillary, then all procedures will produce very similar model checks.

L&L suggest choosing for group $i$ a $T_i$ which is a function of the data $\mathbf{X}_i$ (and possibly the parameters) and as $U_i$ the rest of the data. As L&L indicate, there might be some concern about losing power, but certainly the behavior is much better than that of posterior predictive measures (as clearly shown by L&L's Table 1). As we remarked before, this avoids double use of the data *if we were only testing that group*. Our main concern is how to properly interpret all these $T_i$'s jointly. L&L have been very careful not to compute any $p$-value based on overall measures. For instance, using the overall discrepancy measures $T_1 = \max\{\bar{X}_i\}$, $T_2 = \max\{|\bar{X}_i - \bar{\bar{X}}|\}$ and $T_3 = \max\{|\bar{X}_i - \mu|\}$ produces $p$-values equal to $0.479, 0.619$ and $0.476$, respectively, thus showing the same undesirable behavior as posterior predictive $p$-values, and the concern about double use of the data still arises. (For a simple example of similar issues with cross-validation $p$-values, see the rejoinder to Professor Carlin in Bayarri and Berger (1999).) If we keep the $p$-values individually, it is not very clear what to do with them. One concern is that they are probably highly correlated, and then displays of uniformity might mean little; another important concern is with multiplicity issues, especially when there are many groups. Of course the multiplicity issue gets worsened when, in addition to having many groups, one considers many $T$'s for each group. The only way that we know to satisfactorily handle multiplicities is Bayesian model selection analysis, and the complexity of the problem escalates (and again requires proper priors).

## METHODOLOGICAL ISSUES

In the discussion, various interesting methodological issues arose. We briefly address the main issues here.

*Model elaboration.* Gelman and Johnson touch on model elaboration followed by inference as an alternative to model checking. In the situation contemplated in this paper, however, in which we are seriously entertaining a model, an analysis with a single, more complex model would not be adequate. Correct Bayesian analysis should acknowledge the uncertainty in the model assessment, utilizing model selection (between the more elaborated and the simpler models) or model averaging. This is indeed the ideal Bayesian analysis, but both the analysis and the prior assessments are considerably harder than those required for our model checking proposal. Avoiding the full model uncertainty analysis in situations where we are reasonably confident in the assessed model was precisely the motivation for developing an objective Bayes model checking procedure. Of course, if the model is found incompatible with the data, then a full model selection analysis cannot be avoided.

*Avoiding double use of the data.* Evans suggests that, to avoid double use of the data, our choices for $T$ and $U$ should satisfy his factorization of the joint distribution, at least asymptotically. There is no need for this: we avoid double use of the data by conditioning. Also, there is no need for $T$ and $U$ to be independent (as when splitting the data), nor for $T$ to be sufficient nor for $U$ to be ancillary (in our notation, not Evans'). Computing a mean and a variance of the same posterior distribution is not using the data twice; it is describing two characteristics of that distribution. Similarly, focusing on one "slice" (a conditional distribution) of the joint prior predictive $m(\mathbf{x})$ is not using the data twice, but using a specific characteristic of that distribution. To illustrate with the simplest discrete example, if $T = (x_1, x_2)$ and $U = x_1$, then $m(t \mid u_{obs}) = m(x_1, x_2 \mid x_1 = u_{obs}) = m(x_2 \mid x_1 = u_{obs})$ if $x_1 = u_{obs}$ and 0 otherwise; $x_1$ and $x_2$ are used for different things, but not used twice. Note that posterior predictive checks cannot be cast in this way. This issue is also discussed at length in the rejoinder of Bayarri and Berger (2000).

*Accounting for uncertainty in the estimates.* Gelman argues that there must be something wrong in our recommendation of plug-in checks over posterior predictive checks, since the former do not account for uncertainty in the estimates. It is true that plug-in checks make two mistakes—using the data twice and ignoring the uncertainty in the estimates—whereas posterior predictive checks only make the first mistake. Crucially, however, the second "mistake" that is made by plug-in checks actually operates in the opposite direction of the first mistake, and brings the resulting $p$-value *closer* to uniformity. This was formally shown to be the case in Robins, van der Vaart and Ventura (2000), but can also be understood intuitively: when the data are very incompatible with the model, posterior predictive (and plug-in) distributions sit in the wrong part of the space (the parameters are overtuned to accommodate for model deficiency) but, since the plug-in distribution is (wrongly) more concentrated than the posterior predictive distribution, it is less compatible with extreme values of test statistics, and hence is less conservative. It is the theorem in Robins, van der Vaart and Ventura (2000) that shows the correction is not an overcompensation, that is, that the plug-in still remains conservative, while possessing more power. The plug-in predictive checks are also often easier to compute. Note that this superior performance of the plug-in checks occurs regardless of the specific form of checking used, that is, whether it is formal or graphical.

## LIMITATIONS

We are sympathetic to the complaints concerning the difficulty of computing partial posterior (and conditional) predictive checks, but it can be done and the difficulty is only in estimating a (usually) univariate density at one point, not a difficult computation compared to most Bayesian computations nowadays. However, we recognize that more work is needed to develop fast and efficient algorithms to carry out the necessary computations. For invariant situations, the computations for posterior simulations from the pivotal quantity are simpler, but only when the computed bounds are satisfactory (and the test procedure adequate); otherwise, proper interpretation of the simulated values (whether for visual displays or numerical computations) requires prior predictive techniques, which not only need a proper prior, but also are of a similar level of complexity as the partial posterior predictive technique. Cross-validation may or may not be simpler to compute. The computations required for $m(g(\mathbf{x}) \mid u)$ for a sufficient statistic $U$ (and $g$ any function of the data) are likely to be formidable; in Bayarri, Castellanos and Morales (2006) we actually suggest use of MCMC computations to generate from $m(g(\mathbf{x}) \mid u)$ which are basically identical to the ones used for conditional and partial posterior predictive distributions. For any $T$ (and discrepancy $D$), Robins (1999) and Robins, van der Vaart and Ventura (2000) suggest how to "center" them so as to produce asymptotically uniform $p$-values, and this can also be a daunting task. Posterior predictive techniques are usually simpler to compute than partial posterior or conditional predictive techniques.

Another limitation of our methodology is that it does not say anything about choosing $T$. Choice of $T$ is equivalent to informally choosing the aspect of the model to be checked. What we advocate, once a statistic $T$ has been chosen to detect incompatibility between data and model, is to locate the observed $t$ in the distribution of $m(t \mid u)$ [or in its approximation $m(t \mid \mathbf{x}_{obs} \setminus t_{obs})$]. In the language of Gelman, one should get the "replicates" for model checks from those distributions. This prescription holds whether $T$ is univariate or multivariate, and whether one uses graphics, residuals, relative surprise, $p$-values or other methods to formally or informally locate $T$ in $m(t \mid u)$. This addresses one of Gelman's concerns. (Of course, if $T$ is multivariate, the definition of the $p$-value is not clear.) We do recognize, however, that choice of $T$ is an important issue. Evans and Johnson have both addressed this issue and their suggestions are certainly sensible and worth considering. We do recommend a specific choice of $U$, namely the conditional MLE. Robert and Rousseau (2002) and Fraser and Rousseau (2005) suggest use of the unconditional MLE instead; this choice is also worth exploring.

## MISUNDERSTANDINGS

In the discussions, a number of the statements made concerning our methodology are incorrect. These statements refer to issues that were discussed in our earlier papers where the methodology was first presented, and so we neglected to review these issues in this paper. We try to straighten out some of these misunderstandings here.

Gelman suggests that our methodology focuses on using $p$-values as a model-rejection rule with specified Type-I errors. This is not the case. We do not fix Type-I errors, nor do we advocate use of $p$-values as formal decision rules (indeed, we are quite opposed to it; see Sellke, Bayarri and Berger (2001), and Hubbard and Bayarri (2003)). Indeed, the methodology is valid whether or not $p$-values are used. We use $p$-values as "measures of surprise": numerical quantifications of the incompatibility of the observed $t$ and the "reference" distribution; another such measure is the relative predictive surprise also explored in the paper (and which can readily be applied to multivariate $T$'s). Alternatively, one can opt for checking informally this incompatibility with graphical displays. The main advantage of $p$-values is pedagogical: statisticians are used to interpreting them. Of course, this familiarity is a detriment when procedures such as posterior predictive $p$-values are used, in that casual users will interpret the $p$-values as arising from a uniform distribution, not suspecting that they are instead arising from a distribution much more concentrated about 1/2.

Gelman and Johnson imply that the methodology can only be applied to simple examples and univariate statistics. This is not so. We use "simple" examples so that the numerical complexity does not obscure the relevant issues. As mentioned earlier, there is nothing in the methodology to prevent it being used with multivariate statistics. Similarly, although we use $p$-values and relative surprise (numerical quantifications), one can use graphical displays of simulations from $m(t \mid u)$ in the same way as the discussants use graphical displays from their proposed distributions.

Johnson conjectures that our $p$-values can be anticonservative. Conditional predictive $p$-values can never be uniformly conservative or anticonservative since, as valid Bayesian $p$-values (i.e., based on the prior predictive distribution), they are uniform *on average.* Partial posterior predictive $p$-values are not only asymptotically equivalent to the conditional predictive $p$-values (for the proposed $u$), but very often they are identical; when they are not, the partial posterior and conditional predictive distributions are extremely similar even after very few observations. Of course, if one has an ancillary statistic, one has exact uniformity, but this is rarely the case.

## CONCLUSIONS

Model checking is subtle and has a variety of aspects, as clearly pointed out by the discussants. Optimal selection of $T$ and $U$ is still an issue, and cross-validation might prove useful. A possible answer is Evans' proposals, but we find them unduly limited. Use of pivotal quantities is certainly a possibility in invariant situations, but proper interpretation in general would ultimately require prior predictive analysis and thus preclude use of improper priors. Techniques that produce $p$-values near 0.5 when the model is obviously wrong are simply bad techniques, whether one uses $p$-values, other characteristics of the reference distributions, or graphical displays. Such techniques can detect truly terrible models, but the fact that they can have such poor detection power means that "passing" such a model check does very little to instill confidence that one has a good model.